\documentclass[apj]{emulateapj}

\usepackage{graphicx}
\usepackage{epsfig}
\usepackage{psfrag}

\begin{document}

\title{Pulsar Timing Array Observations of Massive Black Hole Binaries}
\author{Vincent Corbin and Neil J. Cornish}
\affiliation{Department of Physics, Montana State University, Bozeman, MT
59717}

\begin{abstract}
Pulsar timing is a promising technique for detecting low frequency
sources of gravitational waves. Historically the focus has been on
the detection of diffuse stochastic backgrounds, such as those formed
from the superposition of weak signals from a population of
binary black holes. More recently, attention has turned to
members of the binary population that are nearer and brighter, which stand out
from the crowd and can be individually resolved. Here we show that the timing
data from an array of pulsars can be used to recover the physical
parameters describing an individual black hole binary to good accuracy,
even for moderately strong signals. A novel aspect of our analysis is
that we include the distance to each pulsar as a search parameter, which
allows us to utilize the full gravitational wave signal. This doubles
the signal power, improves the sky location determination by an order
of magnitude, and allows us to extract the mass and the distance to the
black hole binary.
\end{abstract}


\section{introduction}
The detection of gravitational waves via the observation of pulse arrival times from pulsars was first
proposed in the 1970s~\citep{Estabrook:1975,Sazhin:1978,Detweiler:1979wn}. Recent improvements in
pulsar timing have made this method one of the best candidates for the first detection of
gravitational waves. Pulsar timing arrays (PTAs) are sensitive to gravitational wave frequencies
between $f=10^{-9}\,{\rm Hz}$ to $f=10^{-6}\,{\rm Hz}$.  The lower bound is set by the observation time,
30 years being a reasonable cut-off, while the upper bound is set by the sample rate, with
once a week as the goal. There is little motivation to
go to higher sample rates as the PTAs operate in the short wavelength limit - the distance to each pulsar
is larger than the wavelength of the gravitational waves - and the sensitivity falls off as $1/f$ across
the band.

The PTA operating range makes it a excellent tool to search for stochastic backgrounds produced by a
population of slowly evolving supermassive black hole
binaries (MBHBs)~\citep{Rajagopal:1994zj,Wyithe:2002ep,Jaffe:2002rt,Wen:2008hw}. The idea is that
the fluctuating time of arrivals caused by the gravitational waves will
produce a correlated response across the PTA, with a characteristic dependence on the angle between
each pair of pulsars~\citep{Hellings:1983fr}.
Considerable work has gone into producing bounds on the amplitude of the stochastic
background~\citep{Hellings:1983fr,Stinebring:1990px,Lommen:2002je,Jenet:2006sv} and
the development of improved
analysis techniques for future searches~\citep{Jenet:2005pv,Anholm:2008wy,vanHaasteren:2008yh}.
A detection of the black hole stochastic background in the pulsar timing band may improve
rate estimates for binary black hole mergers in the Laser Interferometer Space Antenna (LISA) frequency
band~\citep{Wyithe:2002ep}.

With any population there are always members that are nearer and brighter, and it has been
predicted that several black hole binary systems should be individually resolvable when
the diffuse background is detected~\citep{Sesana:2009}. As a prelude to the first detection,
upper bounds have been placed on the maximum amplitude of individual systems using the
Parkes Pulsar Timing Array~\citep{Yardley:2010kv}. When a detection is made, it has been shown
that an analysis of the timing residuals imparted at the Earth can be used to constrain the
amplitude of the signals to $\sim 30\%$, and the direction to $\Delta\Omega \sim 40 \, {\rm deg}^2$
for a 100 pulsar array and a network signal to noise ratio of ${\rm SNR}=10$~\citep{Sesana:2010mx,Sesana:2010ac}.
But the timing residuals imparted at the Earth only tell part of the story. In addition to the
disturbance at the Earth there is also the disturbance at the pulsar to consider. The pulsar component
of the signal is usually ignored as it depends on the poorly constrained distance to each pulsar. In
cross-correlation studies the pulsar terms average to zero as the projected distance to each pulsar is
different, resulting in a different frequency and phase response to individual binaries.
We show that it is possible to include the pulsar terms in the analysis by enlarging the parameter space
to include the distance to each pulsar in the array. This doubles the
signal power and allows the measurement of the mass and distance to the binary. As an added bonus, the
pointing accuracy improves by an order of magnitude, improving the prospects for finding electromagnetic
counterparts.

The outline of the paper is as follows: Section~\ref{sec:Sig_det_res} describes the response of
the detector to a signal from a black hole binary system. An overview of the Bayesian inference
techniques used to estimate the errors in the parameter recovery is given in Section~\ref{sec:Bayesian}.
Section~\ref{sec:Results} displays and discusses the results. The main results are summarized
and future directions are outlined in Section~\ref{sec:Conclusion}.

\section{Gravitational Waves from Supermassive Black Hole Binaries}
\label{sec:Sig_det_res}

Assuming circular orbits, the orbital velocity of a black hole binary in the PTA frequency band
scales as
\begin{equation}
v \simeq 2.5\times 10^{-2} \left(\frac{f}{10^{-8}Hz}\right)^{1/3}
\left(\frac{M}{10^8M_\odot}\right)^{1/3} \, ,
\end{equation}
where $M=m_1+m_2$ is the total mass, $f$ is the gravitational wave frequency, and we are
using units where $G=c=1$. We conclude from this
that the orbital dynamics is only mildly relativistic, and that the gravitational wave emission is
well described by the lowest order post-Newtonian formulae. Another way of seeing this is to
look at the time to merger, which scales as
\begin{equation}
\label{time_thres}
t_c = 2\times10^{6}\, {\rm years}\left(\frac{10^{-8}\,{\rm Hz}}{f}\right)^{8/3}
\left(\frac{10^8 M_\odot}{\cal M}\right)^{5/3}\, .
\end{equation}
Here ${\cal M}=(m_1m_2)^{3/5}/M^{1/5}$ is the chirp mass, which ranges from $0.44 M$ for equal mass
systems to $0.22 M$ for mass ratios of 1:10. These considerations suggest that the gravitational wave
signal can be modeled as a simple sinusoid of fixed frequency~\citep{Sesana:2010mx,Sesana:2010ac,Yardley:2010kv}.
Allowing for moderate orbital eccentricity introduces the complication of having to consider multiple
sinusoids at harmonics of the orbital period, but the essential picture is unchanged.

A more important effect that was not considered in these earlier studies is the contribution to the
signal from gravitational waves disturbing the pulsars. This introduces a new time-scale into the
problem in the form of the projected Earth-pulsar distance, $d\left(1-\cos(\mu)\right)$, where
$d$ is the distance to the pulsar and $\mu$ is the angle between the line of sight to the pulsar
and the line of sight to the black hole binary. This time-scale is typically of order a few thousand
years. When the pulsar term is included, the effective baseline grows from tens, to tens of
thousands of years (the temporal equivalent of aperture synthesis). The
extended baseline makes it possible to measure the frequency change, and hence, the chirp mass.
The minimum detectable rate of frequency change~\citep{Takahashi:2002ky}
\begin{equation}
\label{minimum_freq}
\dot{f}_{\rm min} \simeq \frac{1}{{\rm SNR}\; T^2}\,,
\end{equation}
sets the minimum chirp mass that can be measured:
\begin{eqnarray}
\label{minimum_chirp}
{\cal M}_{\rm min}&=& 5.8 \times 10^6 M_\odot 
\left(\frac{20}{{\rm SNR}}\right)^{3/5} \times \nonumber \\
&&\left(\frac{10^4 \, {\rm years}}{T}\right)^{6/5}
\left(\frac{10^{-8}\, {\rm Hz}}{f}\right)^{11/5}\, .
\end{eqnarray}
With the chirp mass determined, the amplitude of the signal can be used to solve for the
distance to the black hole binary. The $\mu$ dependence in the effective distance provides
additional information about the sky location of the binary, and this, combined with the
increased SNR from using the full signal, leads to a significant improvement in the
angular resolution of the PTA.

The GW signal from a mildly relativistic black hole binary on a circular orbit is characterized by
eight parameters: the distance to the BH binary $D$; the sky location $\phi$ and $\cos(\theta)$;
the angular frequency $\omega_{o}=\pi f_o$ of the binary orbit when observations begin at Earth;
the orbital inclination $\cos(\iota)$; the orbital phase at the line of nodes $\theta_{n}$; the orientation of
the line of nodes $\phi_{n}$ and the chirp mass ${\cal M}$. The signal can be written as the sum
of two sub-signals~\citep{Hellings:1981xc,Hellings:1982bp}, which we refer to as the pulsar signal
and the Earth signal. The former is due to the disturbance caused by the gravitational wave at
the pulsar, the later to the disturbance at the Earth:
\begin{equation}
\label{superposition}
R(t_{e})=r_{p}(t_{p}) + r_{e}(t_{e}).
\end{equation}
In the plane wave limit, which pertains when $D\gg d$, $t_{p}$ is given by
\begin{equation}
\label{pulsar time}
t_{p}\approx t_{e}-d\left(1-\cos(\mu)\right),
\end{equation}
where $t_e$ is the time at the Earth and $\mu$ is the angle between the line of sight to the
pulsar and the line of sight to the binary. The two parts of the residuals are explicitly given in
\cite{Wahlquist:1987rx,Jenet:2003ew}:
\begin{equation}
\label{residuals}
r(t)=\frac{1}{2\left(1+\cos(\mu)\right)}\left(\hat{a}\otimes\hat{a}\right):\left(r_{+}(t)\mathbf{e}^++r_{\times}(t)\mathbf{e}^\times\right)\,.
\end{equation}
Here $\mathbf{e}^{+,\times}$ are the GW polarization tensors, $\hat{a}$ is the unit vector pointing
from the Earth to the pulsar, and
\begin{eqnarray}
\label{rplus rcross}
r_{+}(t) &=& \alpha(t)\left(A(t)\cos(2\phi)-B(t)\sin(2\phi)\right) \\
r_{\times}(t) &=& \alpha(t)\left(A(t)\sin(2\phi)+B(t)\cos(2\phi)\right)\, ,
\end{eqnarray}
with
\begin{eqnarray}
\label{A B}
A(t) &=& -\frac{1}{2}\sin\left[2\left(\Theta(t)-\Theta_{n}\right)\right]\left[3+\cos(2\iota)\right] \\
B(t) &= &2 \cos\left[2\left(\Theta(t)-\Theta_{n}\right)\right]\cos(\iota).
\end{eqnarray}
The amplitude $\alpha(t)$ can be expressed as:
\begin{equation}
\label{alpha}
\alpha(t)=\frac{{\cal M}^{\frac{5}{3}}}{D\omega(t)^{\frac{1}{3}}}.
\end{equation}
Finally, the orbital frequency and the orbital phase evolve according to
\begin{eqnarray}
\label{phase frequency}
\Theta(t) &= & \Theta_{o}+\frac{1}{32{\cal M}^{\frac{5}{3}}}\left(\omega_{o}^{-\frac{5}{3}}-\omega(t)^{-\frac{5}{3}}\right) \\
\label{orbital_freq}
\omega(t) &=& \left(\omega_{o}^{-\frac{8}{3}}-\frac{256}{5}{\cal M}^{\frac{5}{3}}t\right)^{-\frac{3}{8}} \, .
\end{eqnarray}

\begin{figure}[!ht]
	\begin{center}
  \includegraphics[angle=270,width=0.5\textwidth]{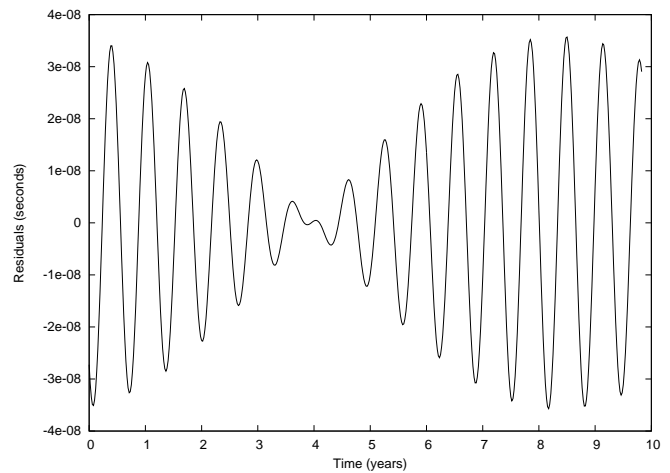}
	\end{center}
\label{R}
\caption{Residuals for a MBHB with chirp mass ${\cal M}=\sqrt{2}\times10^{8}M_\odot$, frequency of the
gravitational perturbation at Earth $f_o=5.0736\times10^{-8}\,{\rm Hz}$ and frequency of the
gravitational perturbation at the pulsar $f=4.7145\times10^{-8}\,{\rm Hz}$}
\end{figure}

Unless the angle $\mu$ is very small, the signal at the pulsar will have a measurably
different frequency
compared to the signal at Earth, since the binary system will have evolved in the elapsed time
$\Delta t=d(1-\cos(\mu))$ (Note: this quantity can be positive or negative depending on who ``sees''
the signal first). Figure \ref{R} shows an example of the noise free timing residuals obtained from a MBHB
system with ${\cal M}=\sqrt{2}\times10^{8}M_\odot$ and initial gravitational wave frequency at the Earth
$f_o=5.0736\times10^{-8}\,{\rm Hz}$. The system is located at a distance of $D=1.77\,{\rm Mpc}$ with a sky location
of $\phi=0.3$ and $\theta=1.42$. The orbital inclination is $\iota = \pi/2$, and the initial orbital phase and
orientation of the line of nodes are both set at $\pi/3$.
The superposition of the two signals is apparent from the beat
envelope. It is possible to recover all the parameters characterizing the MBHB
from the timing residuals
seen in a PTA, so long as the pulsar contributions to the signal are used.

At first sight it may seem impossible to include the pulsar terms in a coherent analysis. To have
phase errors in the pulsar terms less than $\Delta \Theta$ radians requires that we know the distance to
each pulsar to order $\Delta d \sim \Delta \Theta/f$. Setting $\Delta \Theta=0.1$ and $f=10^{-8}{\rm Hz}$
gives $\Delta d \sim  0.1$ parsecs, or a fractional error of $\Delta d/ d \sim 0.01\%$. This is
far smaller than the accuracy that is currently available from electromagnetic observations. Techniques
such as parallax measurements and astrometry achieved a precision of about $10\%$.
Recently~\cite{Verbiest:2008gy} have been able to estimate the distance to a
specific pulsar with $1\%$ error using the kinematic distance
derived from pulsar timing data. This accuracy is not typical, and the method is
only valid for nearby pulsars. But it turns out that highly accurate pulsar distance estimates are
not required if we include the distance to each pulsar, $d_i$, as model parameters to be solved for
from the GW data. The technique works as follows: for any one pulsar the phase matching of the
pulsar terms produces a series of secondary maxima in $d_i$ corresponding to $2\pi$ increments in the
accumulated phase. As the estimate for $d_i$ moves further away from the correct solution along
this line of secondary maxima the predicted GW frequency and amplitude of the pulsar term starts to
deviate from the true value, and the height of the secondary maxima drop. In other words, it is
the overall frequency/amplitude envelope that fixes the distance to the pulsar, and not the phase
matching. Indeed, we will see that the secondary maxima are so close together that the individual
peaks in the posterior distribution for $d_i$ blend together into a continuum. The blending is
even more pronounced in the marginalized posterior distributions for $d_i$ where the correlations
between $d_i$ and $\mu$ are integrated out. We will see that it is possible to use the GW data to
provide estimates for the pulsar distances, while simultaneously deriving useful estimates for the
chirp mass and black hole location.

It was recently pointed out by \cite{Deng:2010ut} that the timing residuals may show departures from
the plane wave approximation used to derive equation (\ref{rplus rcross}). They showed that the curvature of
the wavefronts introduces parallax effects which can be used to measure the distance to the
black hole binary and the pulsars in the array, independent of including the pulsar term in the
signal. Accounting for the wavefront curvature will improve our estimates, 
which neglect this effect.

\begin{figure}[!ht]
	\begin{center}
		\includegraphics[angle=270,width=0.5\textwidth]{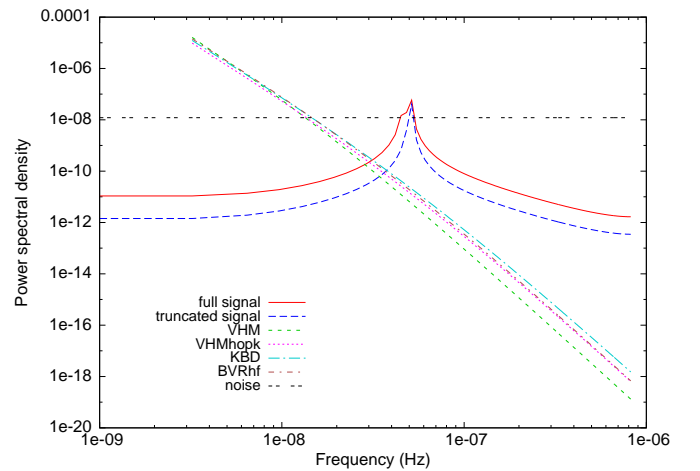}
	\end{center}
\caption{Power spectral density of the full (solid line) and truncated (dashed line) signal from a binary compared with the power spectral density of the instrument white noise and four models of background noise (VHM, VHMhopk, KBD, BVRhf). The signal power
is shown for a single pulsars in the array. The cumulative SNR (over the full array) is 20 for the full signal and 14 for
the truncated signal}
\label{fig:S}
\end{figure}

The common challenge with all GW detectors is to isolate a signal which is relatively weak compared
to the noise surrounding it. The noise is composed of instrumentation or measurement noise and the
background confusion noise generated by the superposition of all the unresolved MBHBs
signals~\citep{Rajagopal:1994zj,Phinney:2001di,Jaffe:2002rt,Jenet:2005pv,Jenet:2006sv,Sesana:2008}.
The measurement noise is simulated as a white noise with a standard deviation of $\sigma=100\,ns$
for all pulsars. An array of 20 pulsars each with a timing rms of about $100\,ns$ is a reasonable
estimate of what will be achieved by PTAs in the near future~\citep{Manchester:2007mx,Verbiest:2009kb}.
It is also thought to be close to the threshold for the detection of the diffuse background~\citep{Jenet:2005pv}.
Models for the background~\citep{Sesana:2008,Sesana:2010mx} suggest that its power spectral density can
be described as:
\begin{equation}
\label{background power spectrum}
S_{\rm bg}(f)=\frac{3H_{o}^{2}}{16\pi}\frac{h_{*}^{2}}{\rho_{c}}f_{*}^{\frac{4}{3}}f^{-\frac{7}{3}}\left(1+\frac{f}{f_{o}}\right)^{2\gamma},
\end{equation}
where $h_{*}$, $f_{*}$ and $\gamma$ are model dependent. This result differs from
earlier studies ({\it e.g.}~\cite{Phinney:2001di}) which predicted a simple $f^{-\frac{7}{3}}$ power-law.
In Figure~\ref{fig:S}, we compare the power spectral density of the signal generated by the MBHB
described above with white instrument noise and four different models of background noise
({\it e.g.}~VHM, VHMhopk, KBD, BVRhf, for a description of these models see \citep{Sesana:2007sh,Volonteri:2002vz,Koushiappas:2003zn,Begelman:2006db,Volonteri:2006qp}).

The square of the signal to noise ratio doubles when using the full signal compared to the truncated signal, which signifies
that including the distances to the pulsars in the parameter space could contribute to the detection of more sources.
For most of the frequency window of interest, the background noise is smaller than both the signal and the
white noise~\citep{Rajagopal:1994zj,Jenet:2005pv,Jenet:2006sv}. For the sake of simplicity, we will ignore the
background for now, as it will not greatly affect the estimation of the parameters for a detected source. One would
need to include it however in order to determine the number of detectable sources.   

\section{Bayesian Inference and Parameter Estimation}
\label{sec:Bayesian}

To determine how well we can estimate the parameters of a MBHB from noisy pulsar timing data
we need to compute the posterior distribution for the model parameters. The
posterior distribution $p(\vec{x}|s)$ gives the probability of observing parameters $\vec{x}$ given
data $s$. It is defined by
 \begin{equation}
\label{posterior distribution}
p(\vec{x}|s)=\frac{p(s|\vec{x})p(\vec{x})}{\int \,p(s|\vec{x})p(\vec{x})\textrm{d}\vec{x}}.
\end{equation}
Here $p(\vec{x})$ is the prior distribution, which is the mathematical representation of
the prior knowledge of the system, and $p(s|\vec{x})$ is the likelihood evaluated at $\vec{x}$.
It is the probability that a system with parameter $\vec{x}$ will yield a signal $s$ in the detector.
 
To generate the posterior distribution, a Markov Chain Monte Carlo (MCMC) algorithm~\citep{Metropolis:1953am,Hastings:1970,Newman,vanHaasteren:2008yh} is used to explore the parameter space. The MCMC consists of proposing ``jumps'' from one location $\vec{x}_{i}$ in the parameter space to another $\vec{x}_{i+1}$. The jumps have a finite probability $\kappa(\vec{x}_{i+1}|\vec{x}_{i})$ of being accepted that is given by the Hasting ratio: 
 \begin{equation}
 \label{acceptance probability}
\kappa(\vec{x}_{i+1}|\vec{x}_{i})={\rm min}(1,H)
\end{equation}
where
\begin{equation}
\label{hasting ratio}
H=\frac{p(\vec{x}_{i+1})p(s|\vec{x}_{i+1})q(\vec{x}_{i}|\vec{x}_{i+1})}{p(\vec{x}_{i})p(s|\vec{x}_{i})q(\vec{x}_{i+1}|\vec{x}_{i})}.
\end{equation}
Here $q(\vec{y}|\vec{x})$ is the proposal distribution: the probability that a jump from $\vec{x}$ to $\vec{y}$ will be proposed. For a more detailed description of MCMC techniques see~\cite{Gamerman}.

\subsection{Prior distribution}

Some information is known prior to the analysis. For instance, there is a higher probability the source
will be located far from the Earth, as the area of a sphere increases as the square of
its radius. To account for this, the prior distribution on $D$ is chosen to be proportional to $D^2$ out
to some large distance $D_{\rm max}$ that is well outside the horizon for PTA detections.
The sources are assumed to be uniformly distributed on the sky, with random orientations.
A more informative prior on the distance and sky location could be built using
galaxy catalogs. For the distances to the $N$ pulsars, the prior follows from
electromagnetic observations\footnote{Paul Demorest has pointed out to us
that for distance estimates derived from parallax measurements, it would be better to choose a
prior that is Gaussian in the parallax $1/d_i$. However,
if the fractional error in $d_i$ or $1/d_i$ is small, the difference between using distance and parallax
to define the prior will also be small. For a recent discussion of the statistics of parallax measurements
see \cite{Verbiest:2010tu}.}, which
we take to be a collection of Gaussians centered around the measured value $d^{EM}_i$
with a standard deviation $\sigma_i=\alpha d^{EM}_i$:
\begin{equation}
\label{prior}
p(d_i)\, \propto \, \displaystyle\prod_{i=1}^{N}e^{-\frac{\left(d^{EM}_i-d_{i}\right)^{2}}{2{\sigma_i}^{2}}}, 
\end{equation}
where $d_{i}$ is the proposed distance to the pulsar $i$. For our simulated PTA we draw $d$ from the
range $0.5-1.5\,{\rm kpc}$, and include errors in the estimated $d^{\rm EM}$ that are consistent with our
choice of prior. We will consider some examples where the fractional distanced error $\alpha = 0.1$
for all the pulsars, and other examples where $\alpha$ takes different values for different pulsars.
We will focus on arrays with $N=20$ pulsars.

\subsection{Likelihood}

The likelihood $p(s|\vec{x})$ is by definition the probability of observing a signal $s$ from a source
with parameters $\vec{x}$. For Gaussian noise it is given by:
\begin{equation}
\label{likelihood}
p(s|\vec{x})=C\exp\left[-\frac{1}{2}\left(\left(s-R(\vec{x})\right)|\left(s-R(\vec{x})\right)\right)\right],
\end{equation}
where $C$ is a normalization constant, $R(\vec{x})$ is the waveform described in Section~\ref{sec:Sig_det_res}
for a set of parameters $\vec{x}$, and the noise weighted inner product is defined as:
\begin{equation}
\label{inner product}
(a|b)=2\int^{\infty}_{0}\left(\tilde{a}^{*}_{i}(f)\tilde{b}_{j}(f)+\tilde{a}_{i}(f)\tilde{b}_{j}^{*}(f)\right){S_n}^{-1}_{ij}(f)df.
\end{equation}
A summation over the indexes $i$ and $j$ from 1 to $N$ is implied. ${S_n}_{ij}(f)$ is the spectral density of the noise correlation matrix $C_{ij}(\tau)=\int n_{i}(t) n_{j}(t+\tau)dt$ where $n_{i}(t)$ is the noise in the signal from pulsar $i$. ${S_n}_{ij}$ is not typically diagonal since the background noise is correlated between the pulsars. However the instrument noise is uncorrelated. Since we ignore the background noise, ${S_n}_{ij}$ becomes diagonal.

\subsection{Proposal distribution}

A combination of six proposal distributions $q(\vec{y}|\vec{x})$ is used. The first makes use of the
Fisher Information Matrix approximation to the posterior distribution.
The Fisher matrix indicates the level of correlation between the parameters, and the
diagonal elements of its inverse give a rough approximation of the error expected in
the estimation of each parameter. The Fisher matrix is the expectation value of the
negative Hessian of the log posterior evaluated at the posterior mode:
\begin{equation}
\label{fisher}
\Gamma_{ij}=-\langle\partial_i\partial_j \log\,p\left(\vec{x}|s\right)\rangle=\left(\partial_iR|\partial_jR\right)-\partial_i\partial_j\log\,p\left(\vec{x}\right)
\end{equation}
In the current setting the prior only contributes to the diagonal elements of the Fisher matrix, adding a term $1/\sigma_i^2$ to the elements representing the pulsars distances. The eigenvectors of the Fisher matrix define a new set of uncorrelated parameters. The eigenvalues indicate the curvature of the likelihood surface along each eigenvector. When the curvature is high, the likelihood changes a lot for a small variation of the ``eigen-parameter'', and a big jump is unlikely to get accepted. To effectively explore the likelihood surface, jumps along those eigenvectors are proposed, which are scaled by their eigenvalues. The second proposal distribution is similar, but with a bigger scaling. Since the jumps are bigger they are less likely to be accepted, but when accepted they explore the parameter space faster. The third proposal consists of drawing a new parameter set from the prior distributions. The fourth proposal pick selects one of the pulsars and draws
a new pulsar distance from the prior distribution. The fifth proposal distribution is similar to
the fourth, but uses a proposal centered around the true value. It prevents the MCMC from
searching exclusively around the values predicted by the prior.  Finally, tiny jumps
are sometimes proposed along each parameter. These tiny jumps are very likely to be accepted,
and this helps prevent the chain from getting stuck in a location where the Fisher matrix
estimates are particularly poor approximations to the posterior distribution.

\subsection{Parallel tempering}
A parallel tempering scheme~\citep{PhysRevLett.57.2607} is implemented
to improve mixing and convergence of the Markov Chains.
It consists of running
a number of chains in parallel, each of them with a ``temperature'' $T=1/\beta$
which modifies the likelihood:
\begin{equation}
\label{temperature}
p_\beta(s|\vec{x}, \beta)= p(s|\vec{x})^\beta.
\end{equation}
The temperature effectively ``smoothes'' the likelihood map. The higher the temperature, the
more likely a jump is going to be accepted. The chains then communicate by swapping with
each other. The swaps have a probability of being accepted given by a Hasting ratio:
\begin{equation}
\label{swaphasting}
H=\frac{p(s|\vec{x}_i, \beta_{j})p(s|\vec{x}_{j}, \beta_{i})} {p(s|\vec{x}_i,  \beta_{i})p(s|\vec{x}_{j},  \beta_{j})} \, .
\end{equation}
The indices $i$ and $j$ refer to the individual chains. Here 10 chains are used, with
temperatures exponentially spaced between $T=1$ and $T=10$.

\section{Results}
\label{sec:Results}

The addition of the pulsar distances to the parameter space allows us to use the pulsar signals in the analysis. It follows that information about the evolution of the orbital frequency over a long period of time and the relative phases between the Earth signal and the pulsar signals can be extracted. This information is enough to independently determine the distance to the binary and the chirp mass. 

\begin{figure}[!ht]
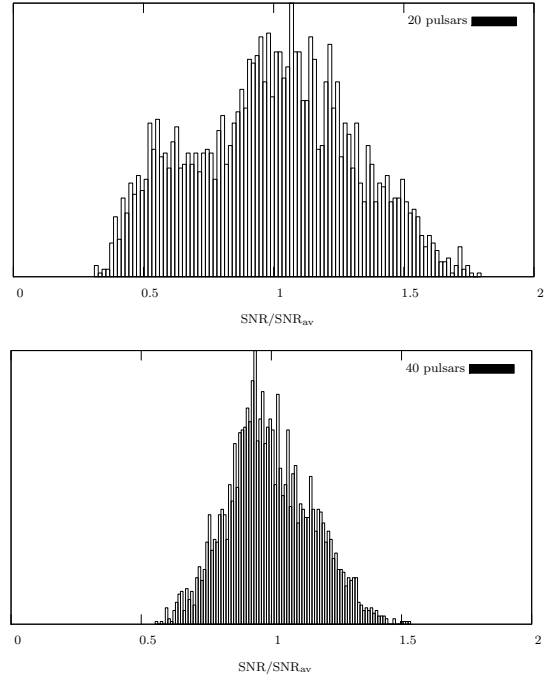

\begin{center}
\scalebox{0.6}{
\begin{tabular}{c}
\input{SNRdist20.txt}\\
\input{SNRdist40.txt}
\end{tabular}
}
\caption{Typical SNR spread across the sky for MBHB signals observed by arrays of 20 and 40 pulsars}
\label{fig:SNR_location}
\end{center}
\end{figure}

We use a simulated PTA comprised of $N=20$ randomly chosen pulsars drawn uniformly in sky location and
with distances in the range $0.5-1.5\,{\rm kpc}$. The parameters describing the array are listed
in Table~\ref{table:pulsars}. With 20 uniformly distributed pulsars the array has a fairly uniform
antenna pattern. Increasing the size of the array to 40 pulsars would further improve the sky
coverage, but the effect is not very significant. To investigate these effects we studied the
the spread in SNR across the sky for a particular source, by considering 3072 different sky locations
using 50 randomly drawn arrays of pulsars. In each case the SNR corresponding to the sky location
was calculated, and normalized by the average SNR to produce the histograms seen in
Figure~\ref{fig:SNR_location}. The variation in SNR across the sky is smaller for the larger PTA,
but not by a significant amount. In either case, we do no expect to see a significant correlation between
the sky location and black hole distance parameters, and this expectation is borne out in our
analysis. The distance and sky location of the 20 pulsars used in our analysis are listed in
Table~\ref{table:pulsars}. 

\begin{table}[!ht]
\begin{center}
\scalebox{0.8}{
\begin{tabular}{|l||c|c|c|c|}
	\hline
	&$d\,(kpc)$&$d_{EM}\,(kpc)$&$\phi$&$\cos\,\theta$ \\
	\hline \hline
	${\rm pulsar}\,\, 1$&$0.987$&$1.010$&$1.17$&$-0.27$\\
	\hline
	${\rm pulsar}\,\, 2$&$1.391$&$1.371$&$5.49$&$0.95$ \\
	\hline
	${\rm pulsar}\,\, 3$&$1.008$&$1.299$&$6.28$&$-0.25$ \\
	\hline
	${\rm pulsar}\,\, 4$&$1.128$&$1.057$&$5.21$&$-0.99$ \\
	\hline
	${\rm pulsar}\,\, 5$&$1.075$&$1.111$&$5.87$&$0.63$ \\
	\hline
	${\rm pulsar}\,\, 6$&$0.651$&$0.629$&$5.23$&$-0.48$ \\
	\hline
	${\rm pulsar}\,\, 7$&$0.957$&$0.895$&$0.02$&$0.67$ \\
	\hline
	${\rm pulsar}\,\, 8$&$1.138$&$0.955$&$5.13$&$-0.37$ \\
	\hline
	${\rm pulsar}\,\, 9$&$1.133$&$0.995$&$3.24$&$0.10$ \\
	\hline
	${\rm pulsar}\,\, 10$&$0.516$&$0.771$&$5.23$&$-0.72$ \\
	\hline
	${\rm pulsar}\,\, 11$&$1.184$&$1.084$&$4.53$&$0.09$\\
	\hline
	${\rm pulsar}\,\, 12$&$1.674$&$1.491$&$5.85$&$-0.52$ \\
	\hline
	${\rm pulsar}\,\, 13$&$1.455$&$1.437$&$3.57$&$0.48$ \\
	\hline
	${\rm pulsar}\,\, 14$&$0.544$&$0.507$&$2.83$&$-0.70$ \\
	\hline
	${\rm pulsar}\,\, 15$&$0.897$&$0.925$&$0.34$&$-0.28$ \\
	\hline
	${\rm pulsar}\,\, 16$&$0.756$&$0.688$&$4.71$&$-0.98$ \\
	\hline
	${\rm pulsar}\,\, 17$&$1.318$&$1.331$&$6.21$&$0.09$ \\
	\hline
	${\rm pulsar}\,\, 18$&$0.882$&$0.920$&$1.46$&$0.14$ \\
	\hline
	${\rm pulsar}\,\, 19$&$0.676$&$0.641$&$0.23$&$-0.64$ \\
	\hline
        ${\rm pulsar}\,\, 20$&$0.852$&$0.757$&$5.60$&$0.09$ \\
	\hline
\end{tabular}
}
	\caption{The position and distances for the 20 pulsars that define the array
used in the analysis. Here $d$ is the true distance to the pulsar, while $d_{EM }$ is
the prior estimate of the distance from traditional astronomical methods.}
	\label{table:pulsars}
\end{center}
\end{table}

\begin{figure}[!h]
	\begin{center}
  \includegraphics[angle=0,width=0.5\textwidth]{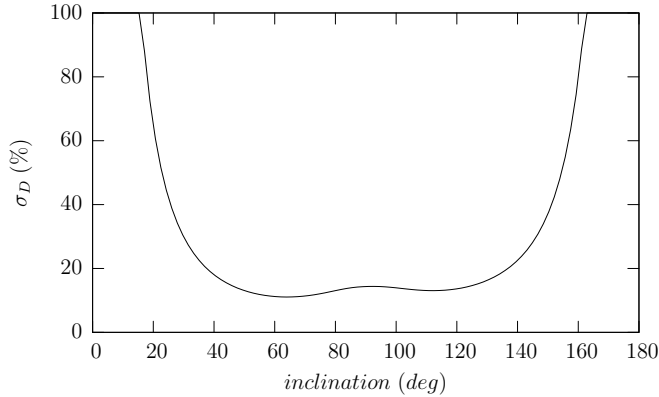}
	\end{center}
	\caption{Error in the determination of the distance to the binary as a function of the inclination.
The error is estimated using the Fisher information matrix. The breaking of reflection symmetry about
$\iota = 90^o$ comes from using a fixed angle to the line of nodes, which means that polarization angle
changes with $\iota$.}
	\label{fig:D_i}
\end{figure}

The distance to the binary is only present in the amplitude of the signal. It makes its determination very dependent on the inclination angle $\iota$. For values of the inclination angle close to $0$ or $\pi$, the determination of the distance to the black hole binary system is very poor. At $\iota=0$ and $\iota=\pi$, the Fisher matrix becomes singular. Figure~\ref{fig:D_i} displays the error in the distance to the binary predicted by the Fisher information matrix as a function of the inclination.

\begin{figure}[!ht]
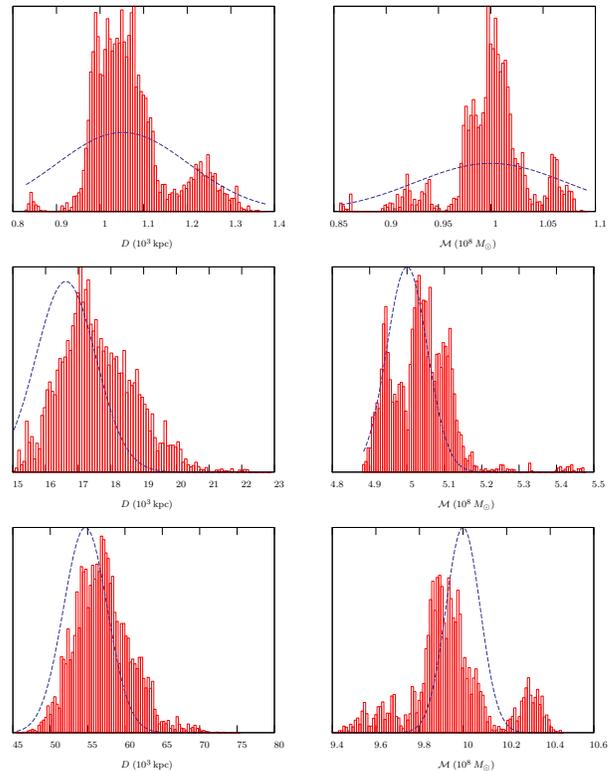

\begin{center}
\scalebox{0.45}{
\begin{tabular}{cc}
\input{1_1_D.txt} & \input{1_1_Mc.txt}\\
\input{1_2_D.txt} & \input{1_2_Mc.txt} \\
\input{1_3_D.txt} & \input{1_3_Mc.txt}
\end{tabular}
}
\caption{Posterior distributions of $D$ and ${\cal M}$ with $\iota=\pi/2$ for three binaries with
different chirp mass: $10^{8}M_{\odot}$ (first row), $5\times10^{8}M_{\odot}$ (second row),
$10^{9}M_{\odot}$ (third row). The distance to the three binary are normalized so that
the three SNR are equal to 20. The MCMC derived posterior distributions (boxes) are compared to
the Fisher matrix estimates (dashed line).}
\end{center}
\label{fig:D_Mc_f_1}
\end{figure}

 For an ideal value of $\pi/2$ for the inclination and a SNR of 20, our MCMC predicted an error of $\sim 7\%$ in the distance estimation, which is in agreement with the Fisher prediction. The error on the chirp mass was $\sim 2\%$. The orbital frequency was extremely well determined ($\sim 0.3\%$). Figure~\ref{fig:D_Mc_f_1} shows the marginalized posterior distribution for the black hole distance and the chirp mass
for three binaries with different chirp masses, $10^{8}M_{\odot}$,$5\times10^{8}M_{\odot}$ and $10^{9}M_{\odot}$ respectively (Sources 1, 2 and 3 in Table~\ref{table:sources}). The marginalized posterior distributions are represented by the boxed histograms
while the smooth line represents the Fisher matrix estimates.
The distances were chosen in order to ensure a SNR of 20 for each case. 
The heaviest source is sufficiently relativistic that higher order post-Newtonian effects may be
detectable, but we defer this analysis to a future study.  
 
The Fisher information matrix is seen to provide a good approximation to the posterior
distribution for the black hole parameters. The differences can in part be attributed to
imperfect convergence in the MCMC runs, which suffer from the high dimensionality of the
parameter space and strong correlations between many of the model parameters.
In some cases the posterior distribution peaks are shifted from their injected  values. This
is caused by the priors for the distances to the pulsars, which are not always peaked close to
their true values, as we will show below.

\begin{figure}[!ht]
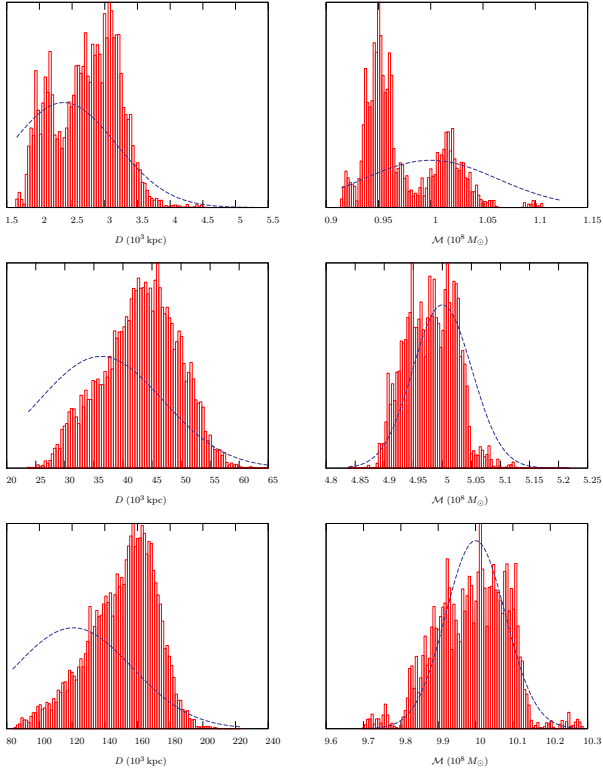

\begin{center}
\scalebox{0.45}{
\begin{tabular}{cc}
\input{2_1_D.txt} & \input{2_1_Mc.txt} \\
\input{2_2_D.txt} & \input{2_2_Mc.txt} \\
\input{2_3_D.txt} & \input{2_3_Mc.txt}
\end{tabular}
}
\caption{Posterior distributions of $D$ and ${\cal M}$ with $\iota=\pi/4$ for three binaries with
different chirp mass: $10^{8}M_{\odot}$ (first row), $5\times10^{8}M_{\odot}$ (second row),
$10^{9}M_{\odot}$ (third row). The distance to the three binary are normalized so that
the three SNR are equal to 20. The MCMC derived posterior distributions (boxes) are compared to
the Fisher matrix estimates (dashed line).}
\end{center}
\label{fig:D_Mc_f_2}
\end{figure}

Unfortunately, most black hole binaries will not have an inclination of $\iota=\pi/2$. For a
more realistic error prediction, we perform the same analysis for similar binaries whose
inclinations are this time chosen to be $\pi/4$. We otherwise use the same sets of
parameters (Sources 4, 5 and 6 in Table~\ref{table:sources}). The distances $D$ change
slightly in order to conserve an SNR of 20. The results are given in Figure~\ref{fig:D_Mc_f_2}.

The error for the distance $D$ is predictably bigger. It rises from less than $10\%$ to
about $15\%$ The Fisher matrix estimates the error to be as large as $30\%$. Again one
has to take into consideration the Fisher matrix gives only a rough approximation to the
posterior distribution. The posterior distributions obtained for the three different binaries are consistent,
which indicates the MCMCs have reasonably converged. The measurements of the frequency and
chirp mass were not significantly affected by the change in
inclination, which confirms they are weakly correlated.

\begin{table}[!ht]
\begin{center}
\scalebox{0.6}{
\begin{tabular}{|c||c|c|c|c|c|c|c|c|}
	\hline
	Source & $D\,(10^3\,kpc)$ & ${\cal M}\,(10^8\,M_\odot)$ & $\omega_o \, (10^{-7}\,s^{-1})$ & $\iota\,(rad)$ & $\phi\,(rad)$ & $\cos\,\theta$ & $\theta_n\,(rad)$ & $\phi_n\,(rad)$ \\
	\hline \hline
	$1$ & $1.05$ & $1.0$ & $1.328$ & $\pi/2$ & $3.98$ & $-0.83070$ & $1.05$ & $6.08$\\
	\hline
	$2$ & $16.6$ & $5.0$ & $1.328$ & $\pi/2$ & $3.98$ & $-0.83070$ & $1.05$ & $6.08$\\
	\hline
	$3$ & $54.7$ & $10$ & $1.328$ & $\pi/2$ & $3.98$ & $-0.83070$ & $1.05$ & $6.08$\\
	\hline
	$4$ & $2.37$ & $1.0$ & $1.328$ & $\pi/4$ & $3.98$ & $-0.83070$ & $1.05$ & $6.08$\\
	\hline
	$5$ & $36.2$ & $5.0$ & $1.328$ & $\pi/4$ & $3.98$ & $-0.83070$ & $1.05$ & $6.08$\\
	\hline
	$6$ & $121$ & $10$ & $1.328$ & $\pi/4$ & $3.98$ & $-0.83070$ & $1.05$ & $6.08$\\
	\hline
\end{tabular}
}
	\caption{The parameters for the 6 black hole binary examples we are considering. The merger times $t_c$
can be computed from these parameters: Sources 1,4 $t_c = 4.39 \times 10^4$ years; Sources 2,5 $t_c = 3.0 \times 10^3$ years;
Sources 3,6 $t_c = 9.46 \times 10^2$ years.}
	\label{table:sources}
\end{center}
\end{table}

\begin{figure}[!ht]
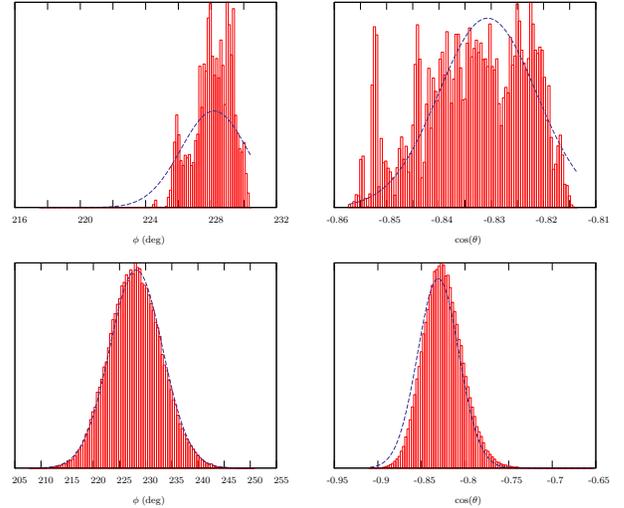

\begin{center}
\scalebox{0.45}{
\begin{tabular}{cc}
\input{2_2_phi.txt} & \input{2_2_cos_theta.txt}  \\
\input{nopulsar_phi.txt} & \input{nopulsar_cos_theta.txt} 
\end{tabular}
}
\caption{Position errors for the full signal (upper panel) and the truncated signal (lower panel).}
\label{fig:position}
\end{center}
\end{figure}

In addition to the decoupling of the binary distance and chirp mass from the amplitude, the addition of the pulsar terms increase considerably the precision of the determination of the binary position. The pulsar terms are evaluated at $t_{p}$, which is the time at which the disturbance from the gravitational wave occurred. It is given by $t_{p}=t-d(1-\cos(\mu))$, where $\mu$ is the angle between the line of sight to the pulsar and the line of sight to the binary. It is therefore a function of the position of the binary. The pulsar signals give previously non-existent information on the chirp mass and distances to the pulsars, but also help refine the measurement of the binary sky location. To illustrate this, Figure~\ref{fig:position} compare the posterior distribution with the Fisher estimation for the two position parameters, for a case in which the full signal was used and a case in which the signal from the disturbance at the pulsar was omitted. When the pulsar signals are added, the SNR naturally increases. Here the distance to the binary was normalized each time to get a SNR of 20 for both cases. The effect from the information encoded in the new term in the signal can then be dissociated from the effect due to the increase in the SNR.  A mid-range chirp mass and frequency
were used (${\cal M}=5\times10^8M_\odot$ and $\omega_o=2\pi/1.5\, {\rm years}$), and an inclination of $\iota=\pi/4$. 

To compare our results to \cite{Sesana:2010ac}, we calculate the pulsar timing array
angular resolution given by
\begin{equation}
\label{angular_resolution}
\Delta\Omega=2\pi\sqrt{(\Delta \cos\theta \Delta\phi)^{2}-(C^{\phi \cos\theta})^{2}}, 
\end{equation}
where $C^{\phi\, \cos\theta}$ is a off-diagonal term of the covariance matrix (the inverse of the Fisher information matrix). For the errors in the individual sky location parameters we use the Fisher matrix estimates, which are in very good agreement with the posterior distributions. For the truncated signal we find $\Delta\Omega=41.3\,{\rm deg}^2$, which is consistent with \cite{Sesana:2010ac}, while the full signal (with the distance changed to preserve the same SNR) yields $\Delta\Omega=6.5\,{\rm deg}^2$. The measurement is therefore improved by a factor of 6.4. Keeping the distance fixed and accounting for the increase in the SNR when
the pulsar terms is included, the angular resolution goes from $\Delta\Omega=41.3\,{\rm deg}^2$ to $\Delta\Omega=2.6\,{\rm deg}^2$. This is a considerable improvement (over an order of magnitude), which highlights
the importance of utilizing the full signal.

It was previously mentioned that the full signal also gives information about the distance to the pulsars. In the simulations
described above we assumed a $10\%$ error in each of the pulsar distances, which correspond to the high end of today's accuracy
in measurements using parallax and other methods. We expect the pulsar distances to be further constrained by the
gravitational wave analysis. The Fisher information matrix predicts that for most pulsars, the error in the
distance measurement will be constrained to a few percent. The level of improvement depends on the location of
the pulsar with respect to the line of sight to the binary. If the Earth , pulsar and binary are aligned, then $t_{p}=t$,
and no new information is given by the pulsar term in the signal. Figure~\ref{fig:pulsar_distance} compares the Fisher
matrix estimates to the MCMC derived posterior distribution and to the prior distribution, for a few relevant pulsars:

\begin{figure}[!ht]
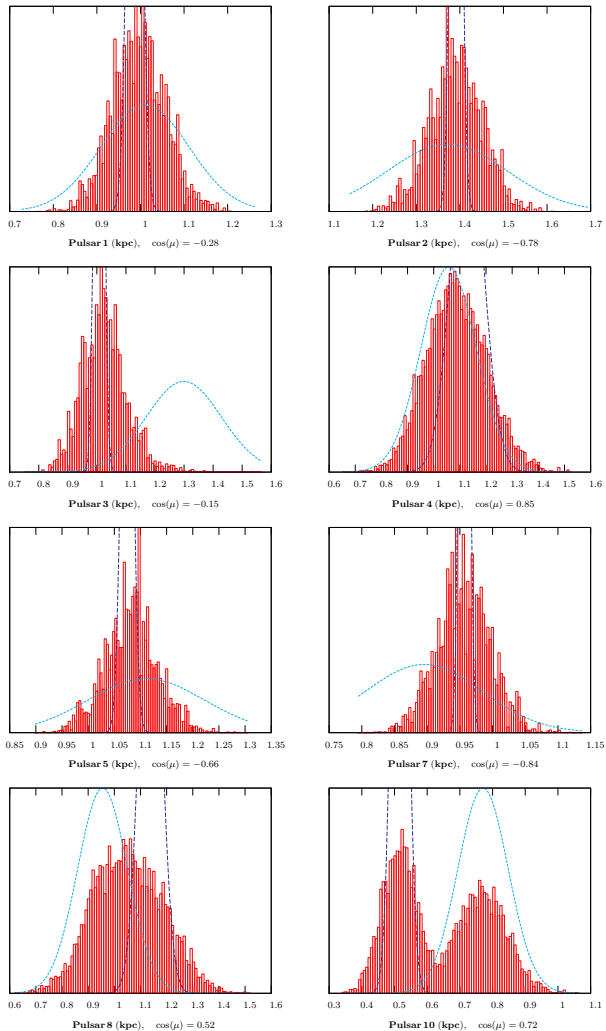

\begin{center}
\scalebox{0.45}{
\begin{tabular}{cc}
\input{2_2_p1.txt} & \input{2_2_p2.txt} \\
\input{2_2_p3.txt} & \input{2_2_p4.txt} \\
\input{2_2_p5.txt} & \input{2_2_p7.txt} \\
\input{2_2_p8.txt} & \input{2_2_p10.txt} 
\end{tabular}
}
\caption{Marginalized posterior distributions for the distances to 8 of the 20 pulsars in the array (boxes) compared to
the prior distributions (dotted line) and the Fisher matrix estimates (dashed line). The distance priors have a
standard deviation of $10\%$, which
correspond to the confidence in the measurement from electromagnetic astronomy. For some of the pulsars, the
gravitational wave signal slightly improves the distance determination.}
\label{fig:pulsar_distance}
\end{center}
\end{figure} 

A few things are clear at first sight. First, when the pulsar's sky location is close to the line of sight to the binary ($\cos\mu\approx\pm1$), the prior is recovered from the posterior distribution, which means that no new information was acquired from the gravitational wave signal, as expected. Then, for the pulsars that are not close to the line of sight, the peak of the posterior distribution can be shifted from the true value of the distance. This occurs when the true value (value at which the Gaussian extracted from the Fisher information matrix is centered) is located a few sigmas away from the prior-predicted value. As mentioned earlier, these shifts will induce a shift in the posterior distribution of the binary distance, chirp mass and orbital frequency with respect to their true values. As a sanity test, the priors were centered to the right values for all the pulsars' distance. As anticipated, the posterior distribution for the binary parameters are in this case centered on their right values as well.
Finally, and maybe most importantly, the posterior distribution from the MCMC is consistently much broader than the error predicted by the Fisher information matrix. For most pulsars, the posterior distribution is as wide as the prior, which means that only limited information concerning the distances could be extracted from the gravitational wave signal. To explain the discrepancy, the pulsar term of the residuals is rewritten as:
\begin{equation}
\label{pulsar_residual}
r_{p}(t)=\frac{{\cal M}^{\frac{5}{3}}}{D\omega^{\frac{1}{3}}(t_{p})}\cos\left(\omega(t_{p})t_{p}+\Phi\right),
\end{equation}
with $t_{p}=t-d(1-\cos\mu)$. If one varies the distance to the pulsar $d$ such that 
\begin{equation}
\label{variation}
\omega(t'_{p})t'_{p}=\omega(t_{p})t_{p}+2\pi,
\end{equation}
then only the amplitude and frequency are slightly changed. This results in a series of secondary maxima
in the likelihood, which for this particular pulsar are spaced by $\Delta d \sim 0.017\,{\rm kpc}$, which
correspond to $1.7\%$ of the distance. The change in frequency and amplitude between adjacent maxima
is small compared to the measurement uncertainty in these quantities, and it is not until multiple
secondary maxima have been traversed that the likelihood drops significantly.
The separation of the secondary maxima is comparable to the Fisher matrix prediction for the error in
$d$, which means that where the Fisher matrix predicts the posterior distribution to drop, the
MCMC will find another maximum, almost as good as the primary. The error in the detection of each
local maximum blends with the error in the detection of its neighbor. They form an ``error envelop'' which
is limited in its width by the change in the frequency and amplitude of the pulsar signal. Even though the
gain in the precision of the measurement of the distances to the pulsars is not as significant as
predicted by the Fisher matrix analysis, it is still noticeable for a few pulsars.

To explore the limitation of the determination of the distances due to the periodicity of the residuals, we set the priors for five pulsars to be Gaussians with standard deviations randomly drawn from the range $\alpha = [0.005,0.03]$. For ten pulsars, the
standard deviation is drawn from the interval  $\alpha = [0.009,0.15]$, and for the remaining five pulsars, the prior
is chosen to be less constraining, $\alpha = [0.20,0.25]$. Table \ref{table:diverse_prior} lists the
estimated distances to the pulsars $d^i_{EM}$, their true distances $d^i$, and the estimated error in the
existing measurement represented by the standard deviation of the prior distribution $\sigma_{\rm prior}$.

\begin{table}[!ht]
\begin{center}
\scalebox{0.8}{
\begin{tabular}{|c||c|c|c|}
	\hline
	Pulsar &$d\,({\rm kpc})$&$d_{EM}\,({\rm kpc})$&$\sigma_{\rm prior}$ \\
	\hline \hline
	$ 1$&$1.009$&$1.010$&$0.53$\\
	\hline
	$ 2$&$1.374$&$1.371$&$1.22$\\
	\hline
	$ 3$&$1.242$&$1.299$&$1.96$\\
	\hline
	$ 4$&$1.070$&$1.057$&$1.75$\\
	\hline
	$ 5$&$1.109$&$1.111$&$0.52$\\
	\hline
	$ 6$&$0.655$&$0.629$&$11.7$\\
	\hline
	$ 7$&$0.965$&$0.895$&$11.2$\\
	\hline
	$ 8$ &$1.139$&$0.955$&$10.0$\\
	\hline
	$ 9$&$1.135$&$0.995$&$10.1$\\
	\hline
	$ 10$&$0.494$&$0.771$&$10.9$\\
	\hline
	$ 11$&$1.184$&$1.084$&$10.0$\\
	\hline
	$ 12$&$1.700$&$1.491$&$11.4$ \\
	\hline
	$ 13$&$1.455$&$1.437$&$10.0$ \\
	\hline
	$ 14$&$0.547$&$0.507$&$10.8$ \\
	\hline
	$ 15$&$0.895$&$0.925$&$10.6$ \\
	\hline
	$ 16$&$0.827$&$0.688$&$20.4$ \\
	\hline
	$ 17$&$1.301$&$1.331$&$23.3$ \\
	\hline
	$ 18$&$0.841$&$0.920$&$20.6$ \\
	\hline
	$ 19$&$0.725$&$0.641$&$24.1$ \\
	\hline
	$ 20$&$0.935$&$0.757$&$18.8$ \\
	\hline
\end{tabular}
}
	\caption{The distances to the 20 pulsars in the array. Here $d$ is the true distance to the pulsar,
while $d_{EM }$ is the prior estimate of the distance by traditional astronomical methods, and $\sigma_{\rm prior}$
is the estimated error in the measurement of the distance (as a percentage of $d_{EM}$).}
	\label{table:diverse_prior}
\end{center}
\end{table}

\begin{figure}[!ht]
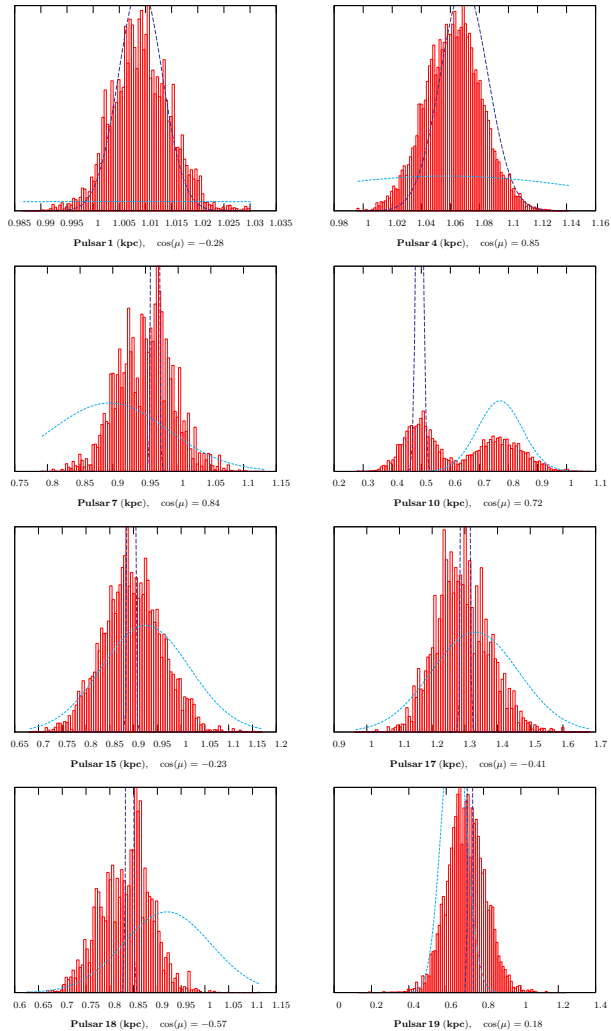

\begin{center}
\scalebox{0.45}{
\begin{tabular}{cc}
\input{diverseprior_p1.txt} & \input{diverseprior_p4.txt} \\
\input{diverseprior_p7.txt} & \input{diverseprior_p10.txt} \\
\input{diverseprior_p15.txt} & \input{diverseprior_p17.txt} \\
\input{diverseprior_p18.txt} & \input{diverseprior_p19.txt} 
\end{tabular}
}
\caption{Marginalized posterior distributions for the distances to 8 of the 20 pulsars in the array (boxes) compared to
the prior distributions (dotted lines) and the Fisher matrix estimates (dashed line). Pulsars $1 \rightarrow 5$ were
assumed to have distances that
were well determined by electromagnetic observations. As a consequence, the measurement of the distances to the
some of the remaining pulsars can be significantly improved by folding in the gravitational wave analysis of
the timing residuals.}
\label{fig:diverse}
\end{center}
\end{figure}

The estimation of the pulsar distance is improved significantly for pulsars 7, 17, 18, 19, 20.
The improvement is largest for pulsars that were poorly constrained by electromagnetic measurements,
and did not lie close to the line of sight to the black hole ($\cos\,\mu\sim 1$). When the prior distance is far
from the real value, the gravitational wave data picks up on the discrepancy, and the posterior distribution becomes
double peaked (pulsar 10). The posterior distribution of pulsar $7$ is identical whether the five first pulsars
have a tight prior or not. This seems to indicate that the improvement on the determination of the distances is
still limited by the periodicity of the strength of the residuals with respect to the pulsars distances. For this reason,
the posterior distribution remains broader than the Fisher estimation. Figure~\ref{fig:diverse} display the posterior
distribution for the distances to a few relevant pulsars against the Fisher estimations and the priors.

The errors on the binary parameters $(D,{\cal M},\omega, \phi, \cos\,\theta)$ were not significantly
affected by the change in the pulsars priors.\\

 \section{Conclusions}
 \label{sec:Conclusion}
We have presented a novel method of analyzing binary black hole signals using pulsar timing data.
By including the distances to the pulsars as model parameters we are able to incorporate the ``pulsar term'' in
the gravitational wave signal, which allows us to detect the decay of the orbit, and hence the chirp mass. This
in turn allows us to convert the amplitude of the signal into a measurement of the distance to the
black hole binary. Including the pulsar terms doubles the signal power and reduces the pointing error by
an order of magnitude. For detections with a network ${\rm SNR}=20$, it should be possible to measure the
distance to $<20\%$, the chirp mass to $<5\%$, and the sky location to $\Delta\Omega < 3\, {\rm deg}^2$.
In something of a role reversal, the gravitational wave observations can improve the distance estimates to
pulsars in the array. The improvement can be significant for pulsars whose distances are originally poorly estimated.
It also follows that the more MBHBs detected, the more trustworthy the estimations for the distances to the pulsars are,
which in turns allow for stronger constraints on the MBHBs detected. 

Topics for future research include the study of systems with eccentric orbits, and the possibility of
detecting higher order post-Newtonian effects (which encode information about
the mass ratio and spin of the black holes) for higher frequency or more
massive systems. It would be interedting to see how including the effects of wavefront
curvature~\citep{Deng:2010ut} improve the distance estimates.
Future studies could also consider the effect of the confusion noise
from the unresolved MBHB background, which will ultimately determine the number of individual systems that
can be resolved.

\section*{Acknowledgments}
We thank Ron Hellings for informative discussions.
This work was supported by NASA Grant NNX10AH15G.

\bibliographystyle{apj}
\bibliography{pulsartiming}	
 
\end{document}